\documentclass[a4paper,11pt]{article}
\usepackage{pos}
\usepackage{lineno}
\usepackage{subfigure}

\title{Latest results of ultra-high-energy cosmic ray measurements with prototypes of the Fluorescence detector Array of Single-pixel Telescopes (FAST)}
\ShortTitle{Latest results of UHECR measurements with prototypes of FAST}

\author*[a,b]{Toshihiro Fujii}
\author[c]{Justin Albury}
\author[c]{Jose Bellido}
\author[d]{Ladislav Chytka}
\author[e]{John Farmer}
\author[d]{Petr Hamal}
\author[f]{Pavel Horvath}
\author[d,f]{Miroslav Hrabovsky}
\author[b]{Hidetoshi Kubo}
\author[f]{Jiri Kvita}
\author[e]{Max Malacari}
\author[d,f]{Dusan Mandat}
\author[g]{Massimo Mastrodicasa}
\author[h]{John Matthews}
\author[d]{Stanislav Michal}
\author[e]{Xiaochen Ni}
\author[b]{Seiya Nozaki}
\author[d]{Libor Nozka}
\author[b]{Tomohiko Oka}
\author[d]{Miroslav Palatka}
\author[d]{Miroslav Pech}
\author[e]{Paolo Privitera}
\author[d]{Petr Schovanek}
\author[g]{Francesco Salamida}
\author[e]{Radomir Smida}
\author[h]{Stan Thomas}
\author[i]{Akimichi Taketa}
\author[b]{Kenta Terauchi}
\author[d,f]{Petr Travnicek}
\author[f]{Martin Vacula}
\author[b]{Seokhyun Yoo}

\affiliation[a]{Hakubi Center for Advanced Research, Kyoto University, Sakyo-ku, Kyoto, Japan}
\affiliation[b]{Graduate School of Science, Kyoto University, Sakyo-ku, Kyoto, Japan}
\affiliation[c]{Department of Physics, University of Adelaide, Adelaide, S.A., Australia}
\affiliation[d]{Institute of Physics of the Academy of Sciences of the Czech Republic, Prague, Czech Republic}
\affiliation[e]{Kavli Institute for Cosmological Physics, University of Chicago, Chicago, IL, USA}
\affiliation[f]{Joint Laboratory of Optics of PU and IF of CAS, Palacky University, Olomouc, Czech Republic}
\affiliation[g]{Department of Physical and Chemical Sciences, University of L'Aquila and INFN LNGS}
\affiliation[h]{High Energy Astrophysics Institute and Department of Physics and Astronomy, University of Utah, Salt Lake City, UT, USA}
\affiliation[i]{Earthquake Research Institute, University of Tokyo, Bunkyo-ku, Tokyo, Japan}

\emailAdd{fujii@cr.scphys.kyoto-u.ac.jp}

\abstract{
  The origin and nature of ultra-high-energy cosmic rays (UHECRs) remain an open question in astroparticle physics. Motivated by the need for an unprecedented aperture for further advancements, the Fluorescence detector Array of Single-pixel Telescopes (FAST) is a prospective next-generation, ground-based UHECR observatory that aims to cover a huge area by deploying a large array of low-cost fluorescence detectors.
  The full-scale FAST prototype consists of four 20\,cm photomultiplier tubes at the focus of a segmented mirror 1.6\,m in diameter.
  Over the last five years, three prototypes have been installed at the Telescope Array Experiment in Utah, USA, and one prototype at the Pierre Auger Observatory in Mendoza, Argentina, commencing remote observation of UHECRs in both hemispheres. We report on the latest results of these FAST prototypes, including telescope calibrations, atmospheric monitoring, ongoing electronics upgrades, development of sophisticated reconstruction methods,
  and UHECR detections.
}

\FullConference{37$^{\rm{th}}$ International Cosmic Ray Conference (ICRC 2021)\\
		July 12th -- 23rd, 2021\\
		Online -- Berlin, Germany}

\begin{document}
\maketitle

\section{Detection of ultra-high-energy cosmic rays}

\label{intro}
Since the discovery of cosmic rays above 100\,EeV ($\equiv$ 10$^{20}$\,eV) in 1963~\cite{Linsley:1963km}, scientists have constructed increasingly-large observatories to detect ultra-high-energy cosmic rays (UHECRs). UHECR sources and acceleration mechanisms at the highest energies are still largely unknown~\cite{AlvesBatista:2019tlv}, making them one of the most intriguing mysteries in particle astrophysics.  Since they are deflected less strongly by magnetic fields (due to their enormous kinetic energies), their arrival directions are more significantly correlated with their sources. Charged-particle astronomy with UHECRs is hence a potentially viable probe of extremely energetic phenomena in the nearby universe.

Two well-established methods are used for UHECR detection: arrays of detectors (e.g. plastic scintillators or water-Cherenkov stations) that sample extensive air shower (EAS) particles at the ground level and large-field-of-view telescopes that directly measure atmospheric shower development by observing ultra-violet nitrogen fluorescence.  The two largest UHECR observatories are hybrid detectors that combine both techniques, employing arrays of ground detectors overlooked by fluorescence detectors (FDs). These are the Pierre Auger Observatory (Auger) in Mendoza, Argentina~\cite{bib:auger}, and the Telescope Array Experiment (TA) in Utah, USA~\cite{bib:tafd, bib:tasd}.

\begin{figure}[b]
  \centering
  \subfigure[Equatorial, Ankle, 45$^{\circ}$ oversampling]{\includegraphics[width=0.49\linewidth]{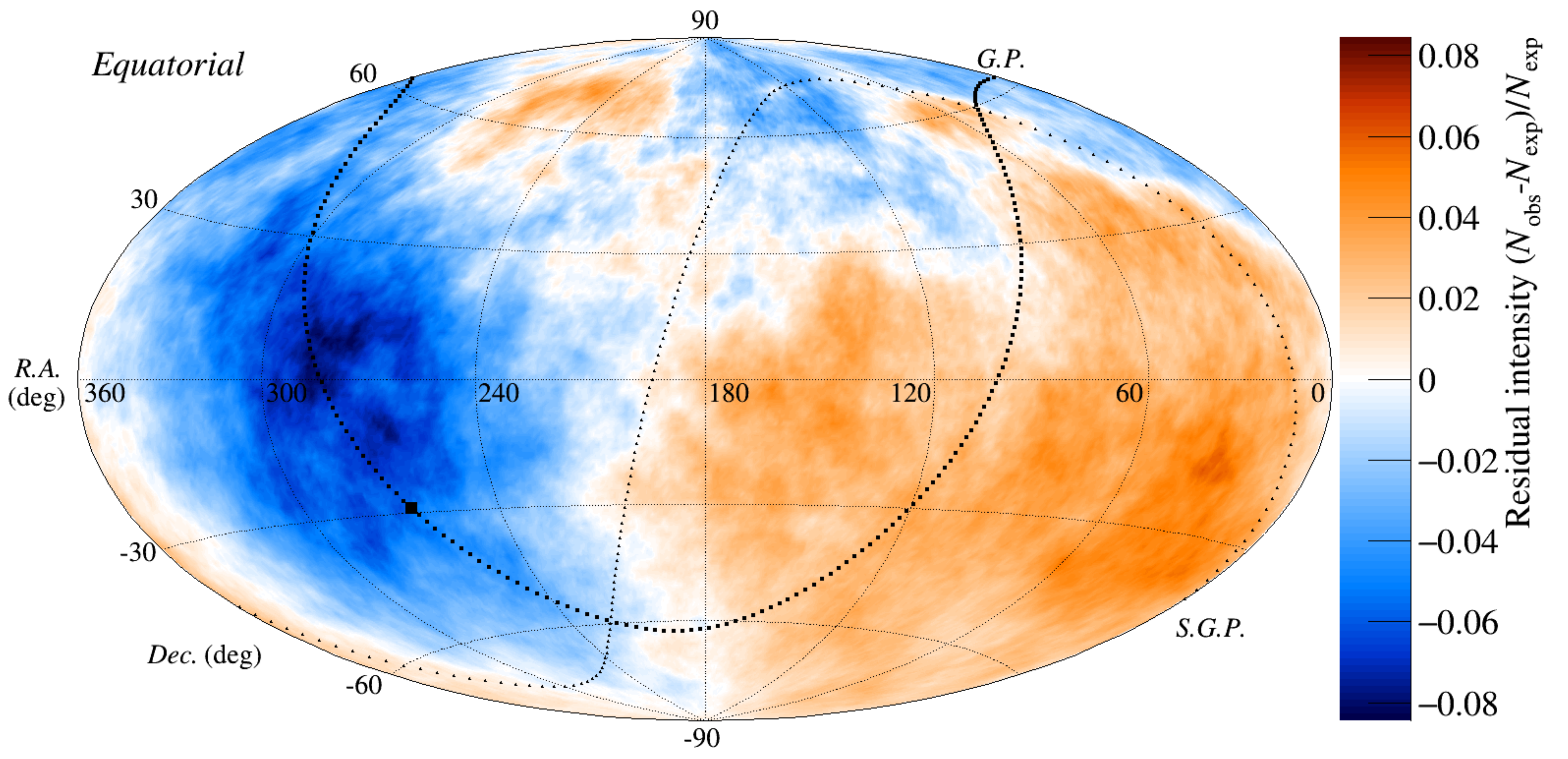}}
  \subfigure[Galactic, Ankle, 45$^{\circ}$ oversampling]{\includegraphics[width=0.49\linewidth]{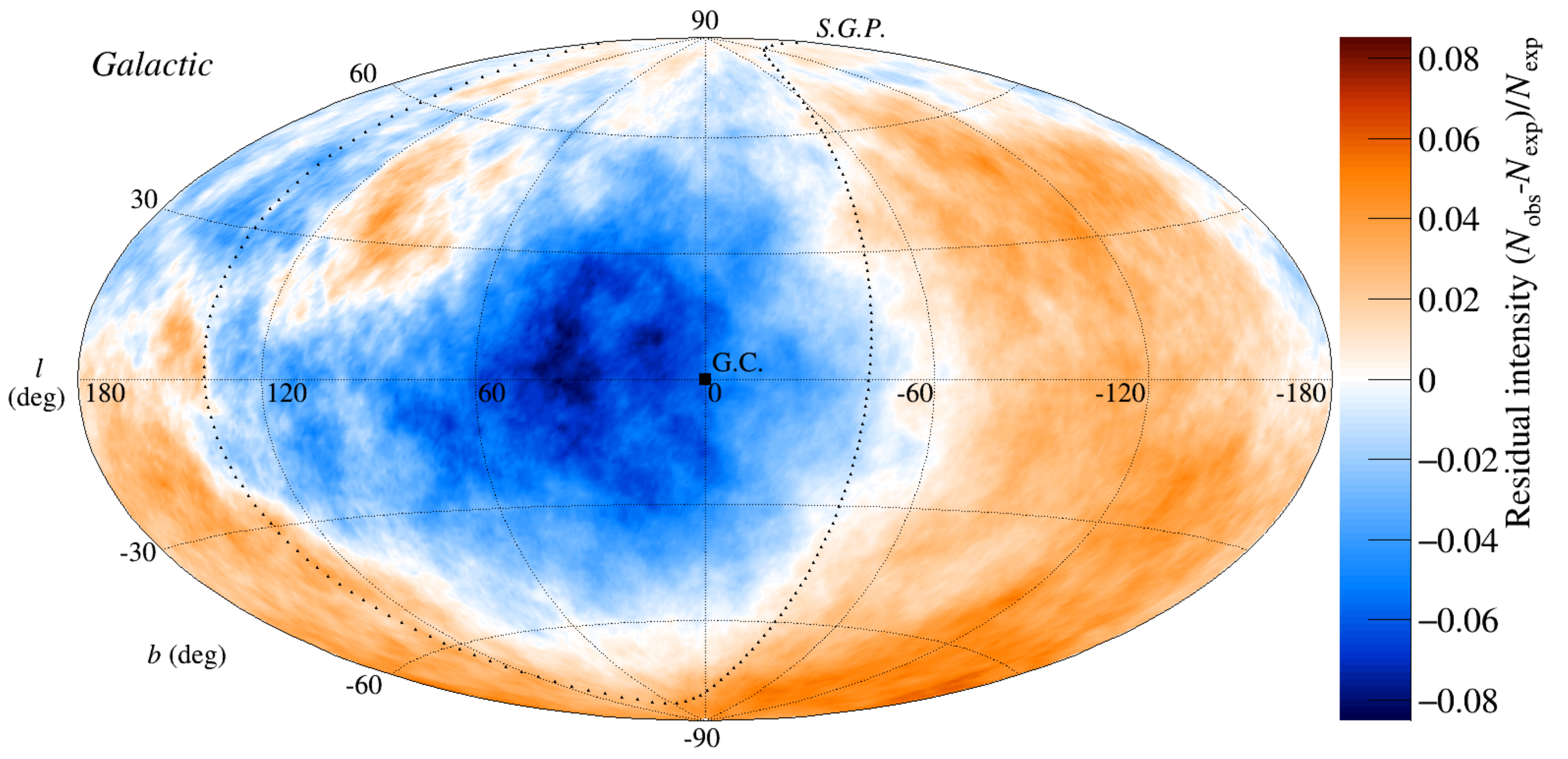}}
  \subfigure[Equatorial, Cutoff, 20$^{\circ}$ oversampling]{\includegraphics[width=0.49\linewidth]{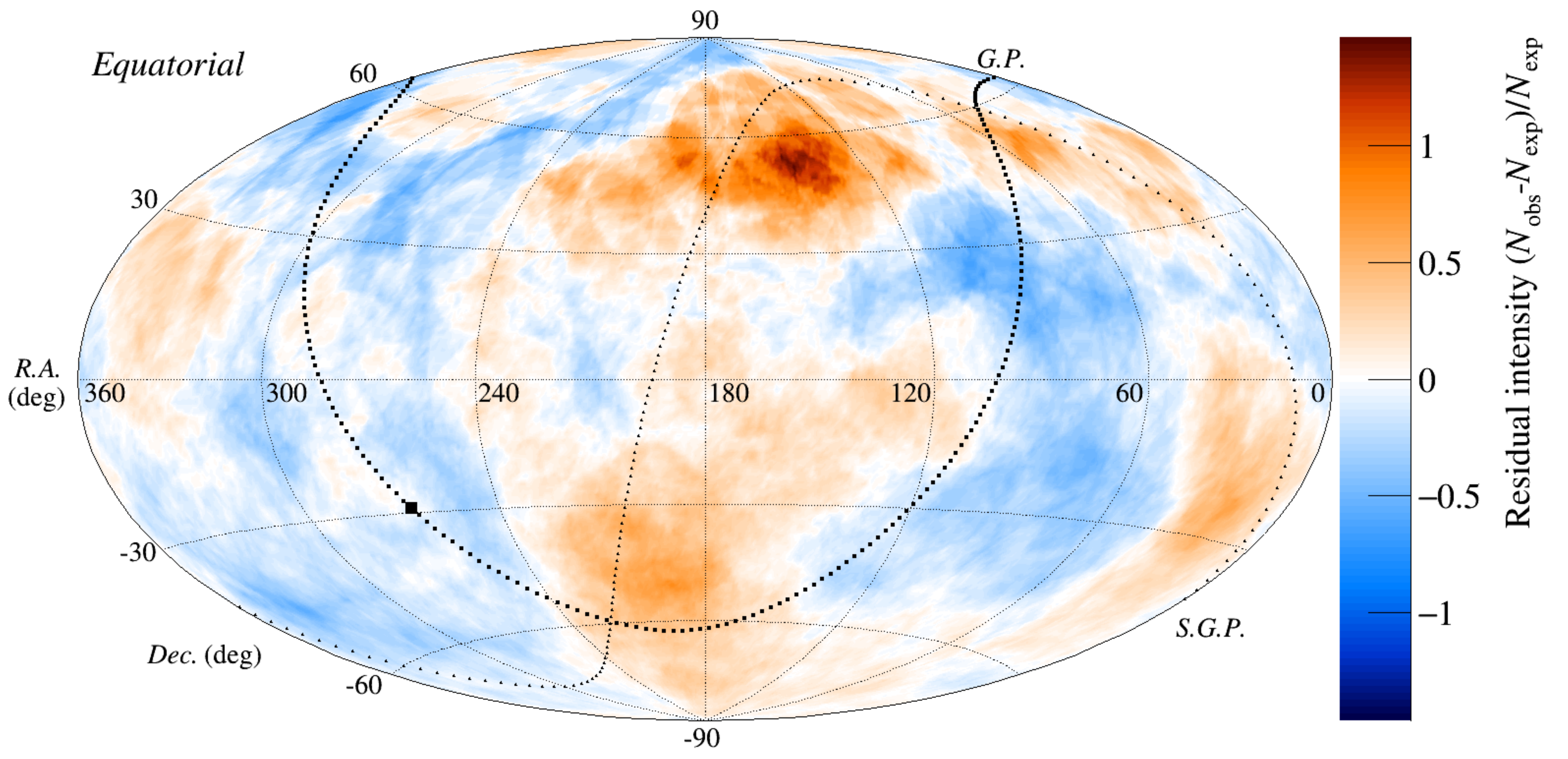}}
  \subfigure[Galactic, Cutoff, 20$^{\circ}$ oversampling]{\includegraphics[width=0.49\linewidth]{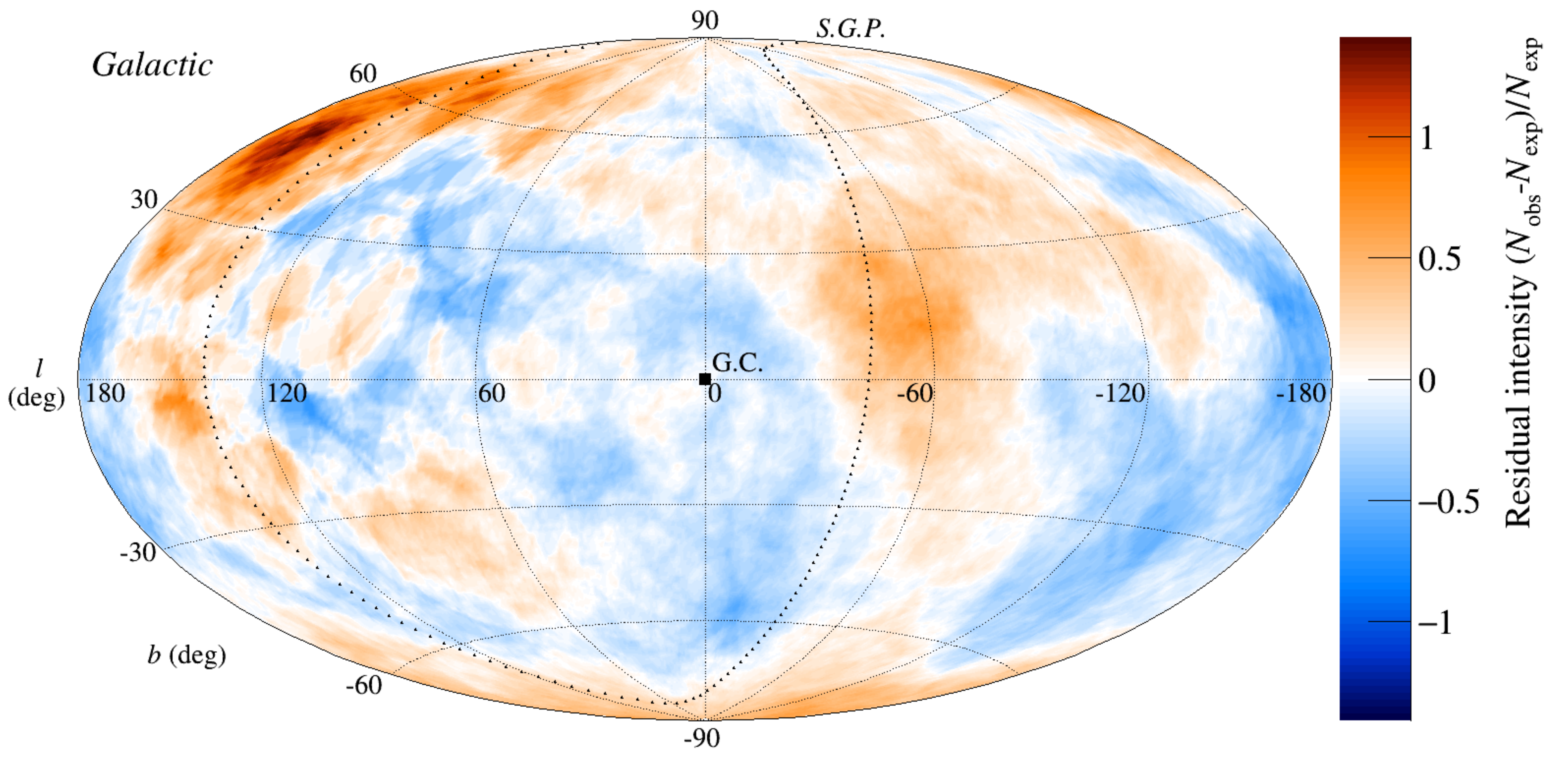}}
\caption{Residual intensity sky-maps in Equatorial and Galactic coordinates:  (a, b) above the ankle energy, with a 45$^{\circ}$ top-hat oversampling; and (c, d) above the cutoff energy with a 20$^{\circ}$ top-hat oversampling. The event data are taken from~\cite{Biteau:2019aaf,diMatteo:2020dlo}.}
\label{bib:skymaps}
\end{figure}

Recent results have shown novel structures at higher energies. Figure~\ref{bib:skymaps} shows full-sky maps of residual intensities measured by both Auger and TA observatories above the ``ankle'' energy of $\sim$10\,EeV and the ``cutoff'' energy of $\sim$50\,EeV, calibrated by an energy scale from the common declination band's flux ~\cite{Biteau:2019aaf,diMatteo:2020dlo}.
Auger reported a large-scale dipole anisotropy above 8\,EeV of 6.5\% amplitude with a 5.2$\sigma$ significance~\cite{bib:dipole_auger}, which supports an extragalactic origin for these particles.
Auger also has reported a 4.0$\sigma$ correlation between the positions of nearby starburst galaxies and the arrival directions of 9.7\% of their measured UHECR events above 39\,EeV~\cite{bib:sbg_auger}.
TA has measured an excess of cosmic rays above 57\,EeV as a ``hotspot'' centered at a right ascension of 147$^{\circ}$ and a declination of 43$^{\circ}$ with a 3.4$\sigma$ significance~\cite{bib:hotspot_ta} and has also reported results consistent with Auger for the dipole search~\cite{Abbasi:2018tqo} and the flux pattern analysis~\cite{Abbasi:2020ohd}.

Further results at the highest energies are limited by statistics due to sharp attenuation of the spectrum. Future ground arrays will require an unprecedented aperture (exceeding current experiments by an order of magnitude) and mass composition sensitivity above 100\,EeV.
Future detectors should hence be low-cost and easy to deploy, operate and maintain. A worldwide collaboration is necessary to construct such an array.

\section{Fluorescence detector Array of Single-pixel Telescopes (FAST)}
\begin{figure}[b]
\centering
\includegraphics[width=1.0\linewidth]{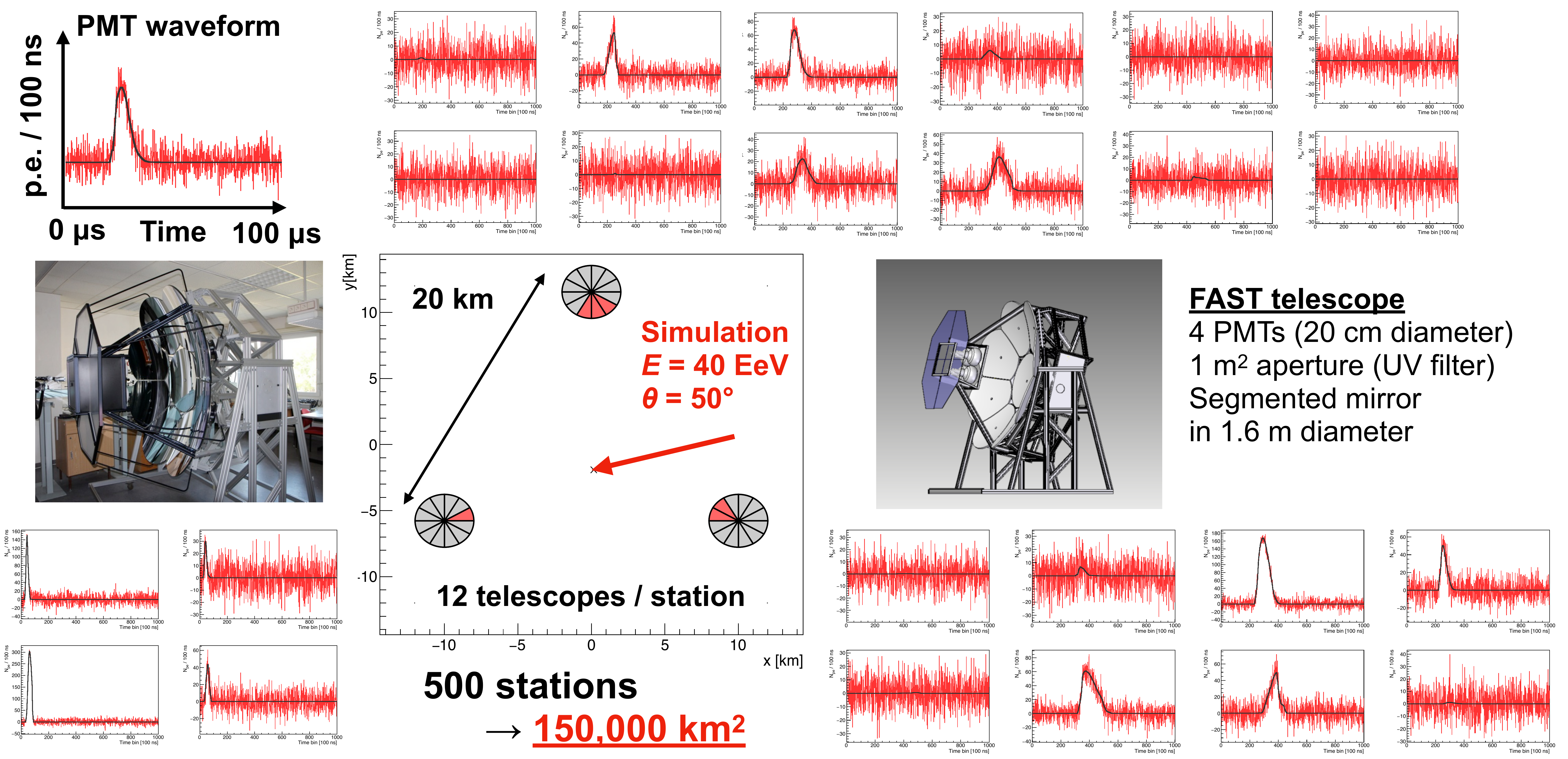}
\caption{The Fluorescence detector Array of Single-pixel Telescopes: a possible solution for a future giant ground array~\cite{bib:fast}. The traces show simulated signals emitted from a UHECR with an energy of 40\,EeV and a zenith of 50$^{\circ}$.}
\label{bib:fast}
\end{figure}
One way to achieve this unprecedented aperture is a ground-based fluorescence detector array.
The Fluorescence detector Array of Single-pixel Telescopes (FAST)\footnote{https://www.fast-project.org} features compact FD telescopes with a smaller light-collecting area and far fewer pixels than current-generation FD designs, leading to a significant reduction in cost that allows for the production of more FD units.

In the FAST design, a 30$^{\circ}$ $\times$ 30$^{\circ}$ field-of-view is covered by four 20\,cm photomultiplier-tubes (PMTs) at the focal plane of a compact segmented mirror of 1.6\,m diameter~\cite{bib:fast_optics}. 
 Its smaller light-collecting optics, smaller telescope housing, and fewer number of PMTs significantly reduces its cost to be $\sim$35\,kUSD per telescope.
Each FAST station would consist of 12 such telescopes, covering 360$^{\circ}$ in azimuth and 30$^{\circ}$ in elevation. These stations would be deployed in a triangular array with a 20\,km spacing, suggested by simulations.
Figure~\ref{bib:fast} shows the simulated waveforms from a UHECR shower detected in 3-fold coincidence by such an array.
To achieve our aperture goals, 500 stations covering 150,000\,km$^2$ are required, after accounting for the standard FD duty-cycle and additional moon-night operation.

\section{Progress of developments on the FAST prototypes}
\begin{figure}[b]
  \centering
  \subfigure[FAST prototypes installed at TA and Auger observatories]{\includegraphics[width=0.8\linewidth]{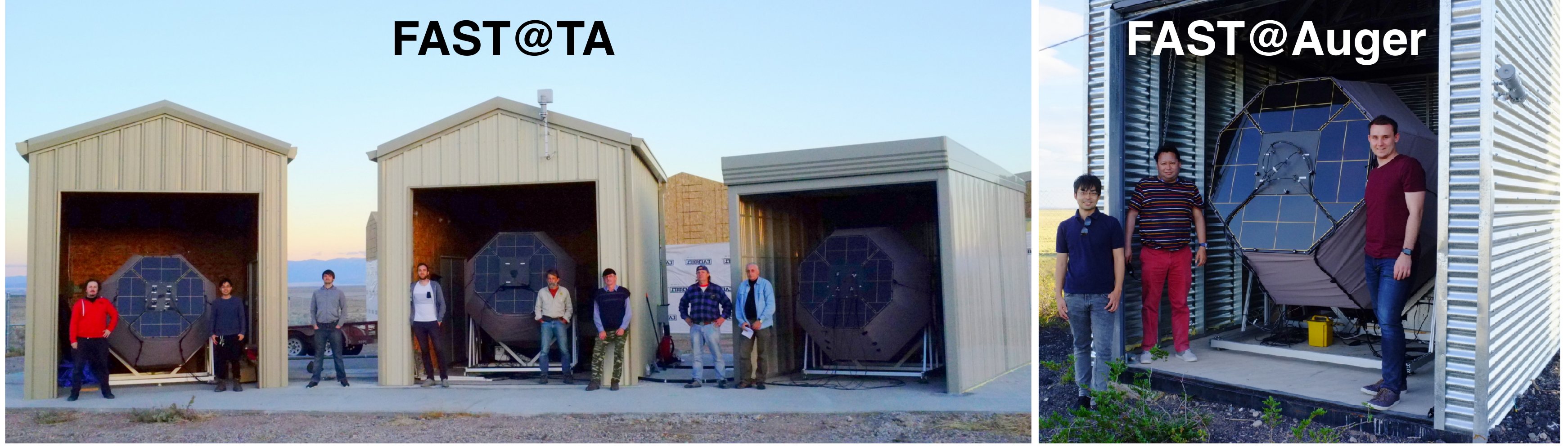}}
  \subfigure[Impact parameter for coincidences between TA FD and FAST]{\includegraphics[width=0.31\linewidth]{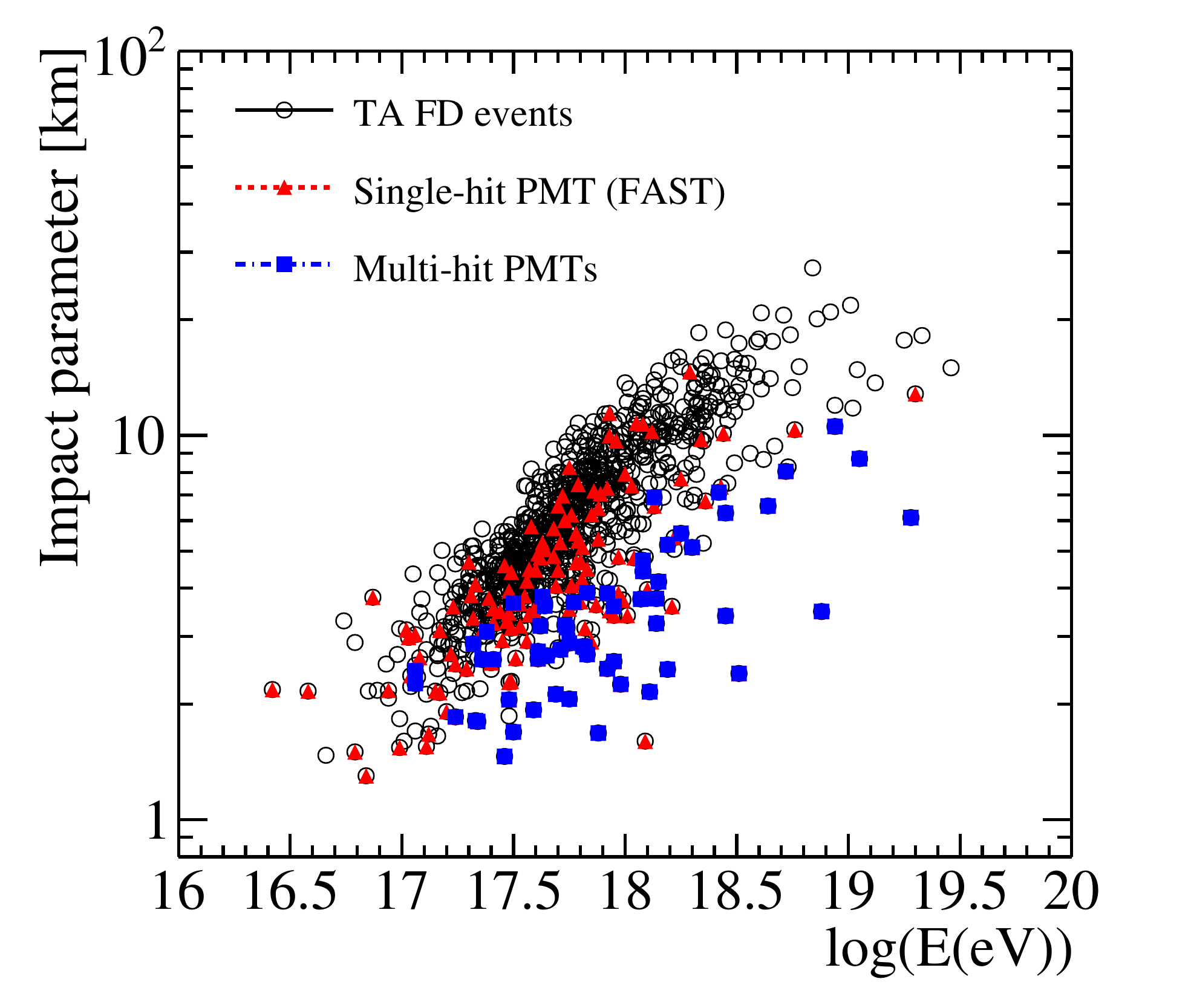}}
\subfigure[Time-average brightness for coincidences between TA FD and FAST]{\includegraphics[width=0.31\linewidth]{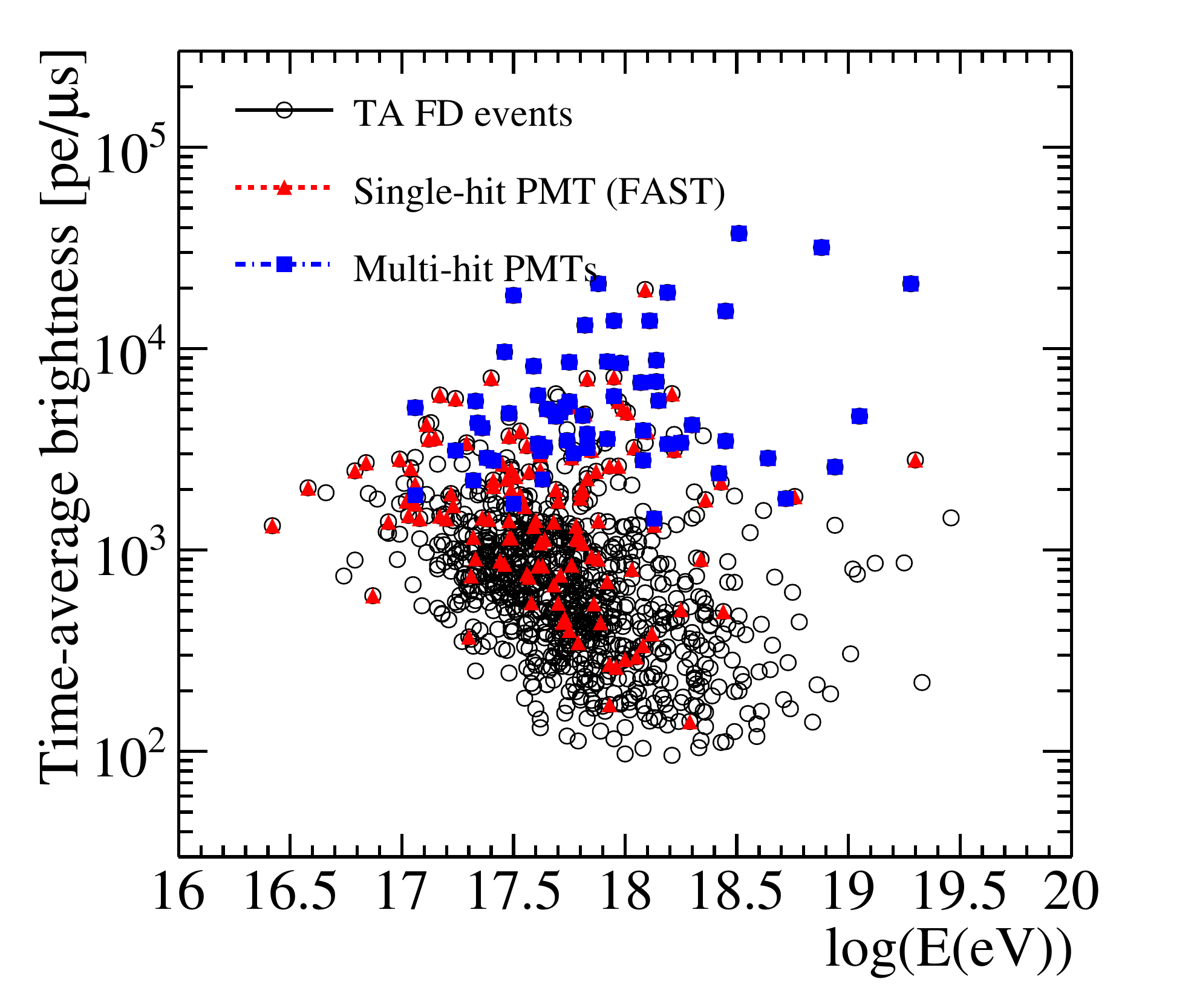}}
  \subfigure[Energy and $X_{\max}$ reconstructed by FAST top-down reconstruction]{\includegraphics[width=0.34\linewidth]{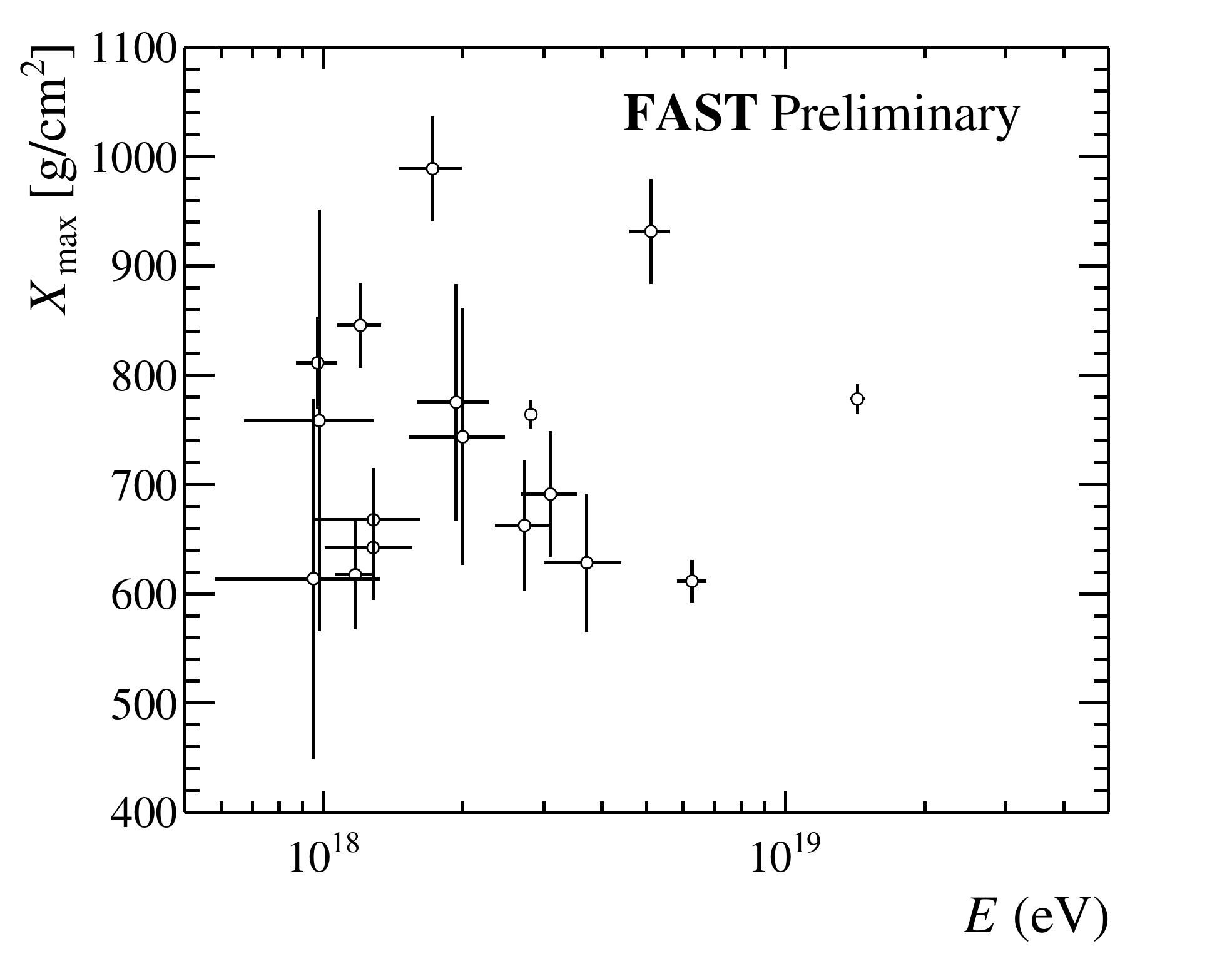}}
  \caption{(a) The three FAST prototypes installed at the Black Rock Mesa site of the Telescope Array Experiment and the one prototype installed at Los Leones site of the Pierre Auger Observatory. (b) Impact parameter and (c) time-average brightness for the coincidence search between TA FD and FAST. (d) Preliminary result of top-down Energy and $X_{\max}$ reconstructions for multi-hit events above 1\,EeV.}
  \label{fig:fast_tel}  
\end{figure}
Motivated by UHECR detections with a single 20\,cm PMT at the focus of a 1\,m$^2$ Fresnel lens in 2014~\cite{bib:fast}, we installed three full-scale FAST prototypes at the TA site from 2016 to 2019, as shown in Figure~\ref{fig:fast_tel}(a)-left~\cite{Malacari:2019uqw}.  We assembled the telescope frames on-site, mounted the PMTs in their camera boxes, and installed ultra-violet band-pass filters at their apertures. 
We then astrometrically aligned the telescopes using a camera mounted to their frames' exteriors~\cite{bib:fast_optics}. Following this, we began observation via remote connection, using external triggers from the adjacent TA fluorescence detector. We used an automated all-sky monitoring camera to record cloud coverage and atmospheric transparency~\cite{Chytka:2020hgv}.
As shown in Figure~\ref{fig:fast_tel}(a)-right, an identical FAST prototype was also installed at the Auger site for a cross-calibration of energy and $X_{\max}$ scales.

Analyzing 224\,hours of data measured by the FAST prototypes at the TA site from March 2018 to October 2019, we found 964 showers with corresponding monocular reconstructions from the TA FD~\cite{bib:tafd_spectrum2016}.
We searched for significant signals (defined as a $\ge 6\sigma$ signal-to-noise ratio over $\ge500$ nanoseconds) in time coincidence with these FD events and found 179 significant FAST events out of the 964 TA EASs, with 59 events producing significant signals in more than one PMT.
Figure~\ref{fig:fast_tel}(b) and (c) show the impact parameter and time-average brightness of the detected EASs as a function of energy, split by single-PMT and multi-PMT events. These parameters are reconstructed by the TA FD.

A ``top-down'' reconstruction algorithm has been implemented that determines the best-fit shower parameters by comparing our measured traces to the simulated ones~\cite{Malacari:2019uqw}. 
Because FAST features only four pixels, rather than use the entry and exit times for each pixel as traditional reconstruction methods do, we extract timing information from each individual bin of the traces.
Figure~\ref{fig:fast_tel}(d) shows preliminary $X_{\max}$ and energy values reconstructed by this method for multi-hit events above 1\,EeV using only FAST prototypes.

\begin{figure}[t]
  \centering
  \subfigure[Reconstructed $\langle X_{\max} \rangle$]{\includegraphics[width=.42\linewidth]{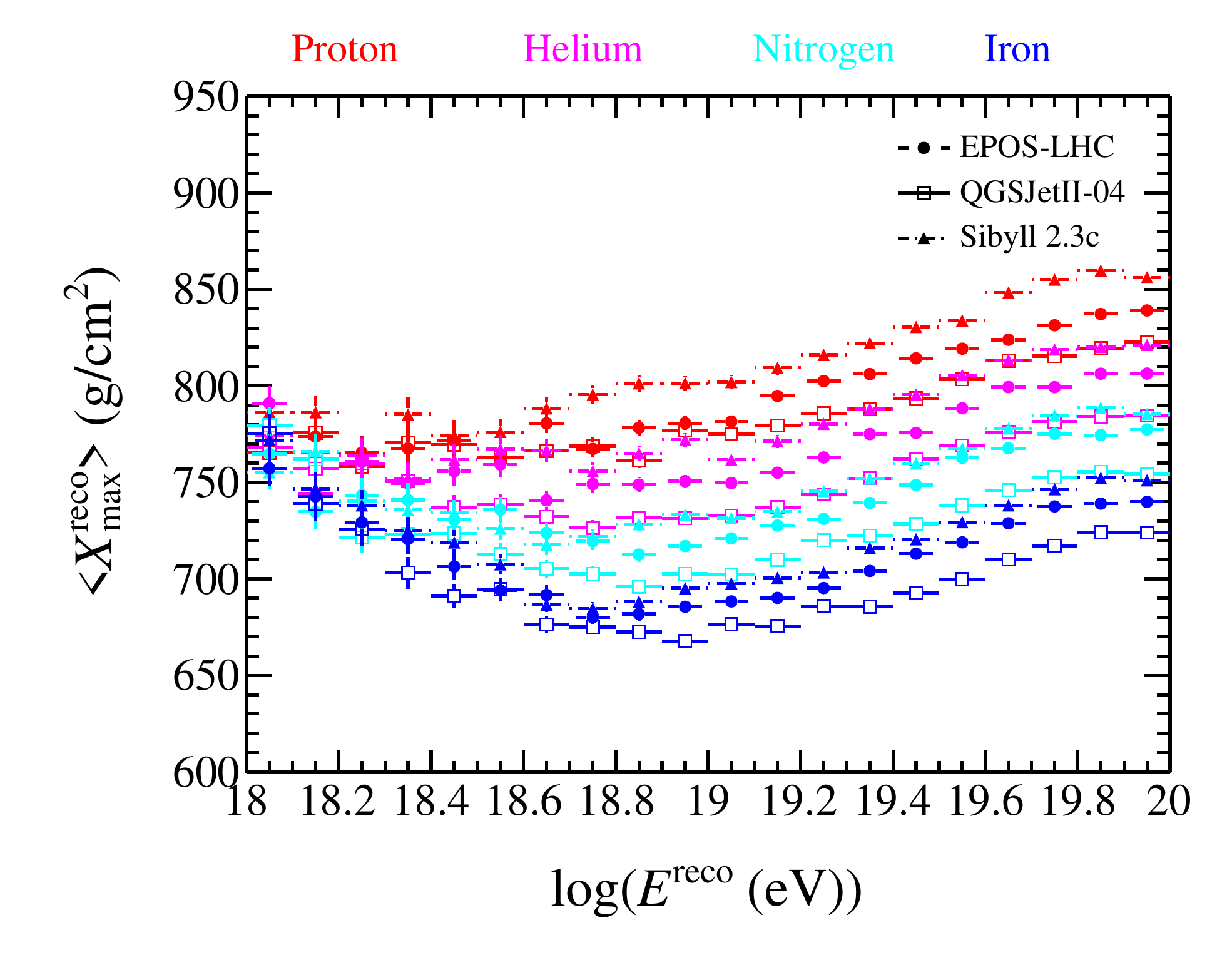}}
  \subfigure[Reconstructed $\sigma(X_{\max})$]{\includegraphics[width=.42\linewidth]{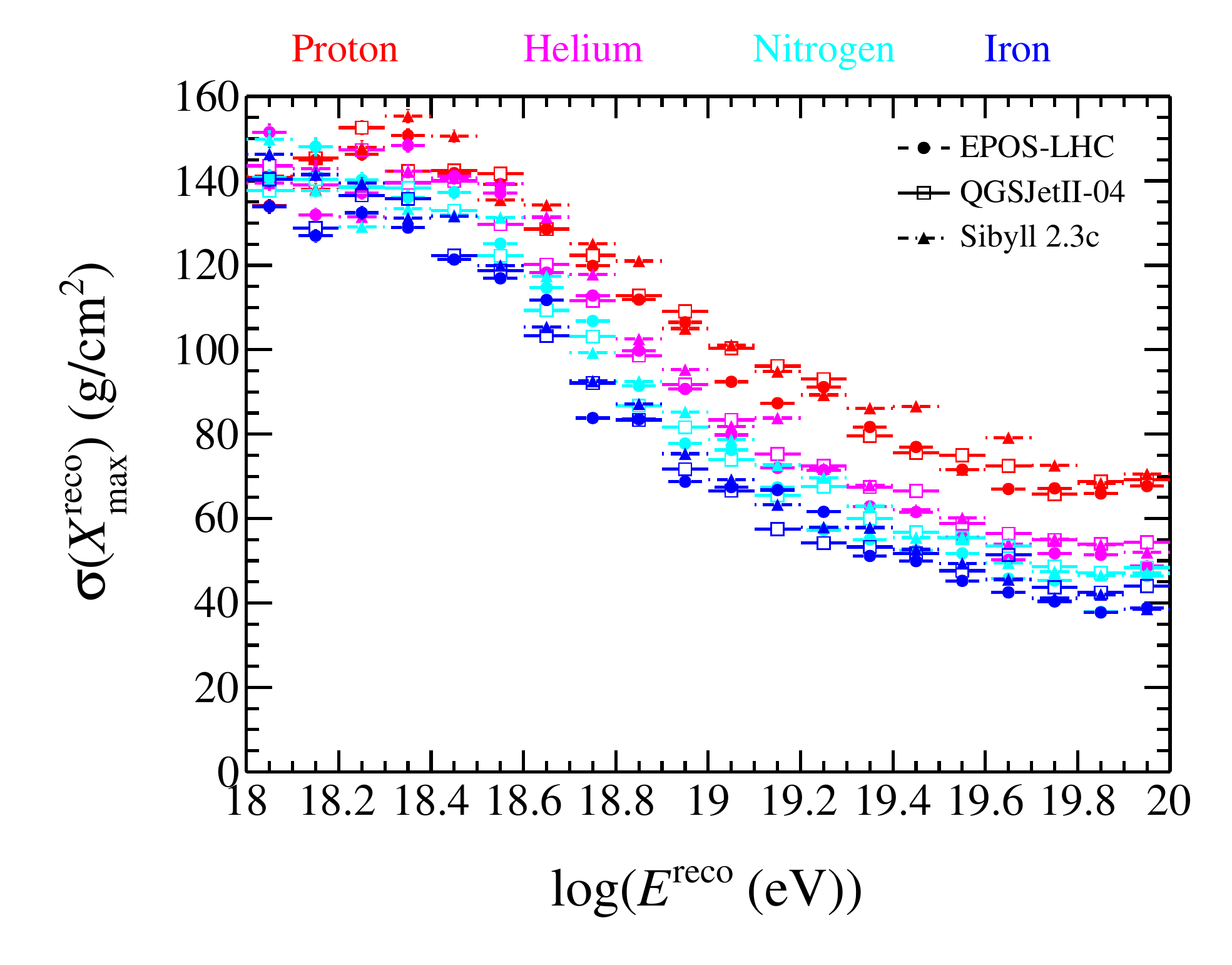}}
  \subfigure[Reconstructed $X_{\max}$ distributions]{\includegraphics[width=0.9\linewidth]{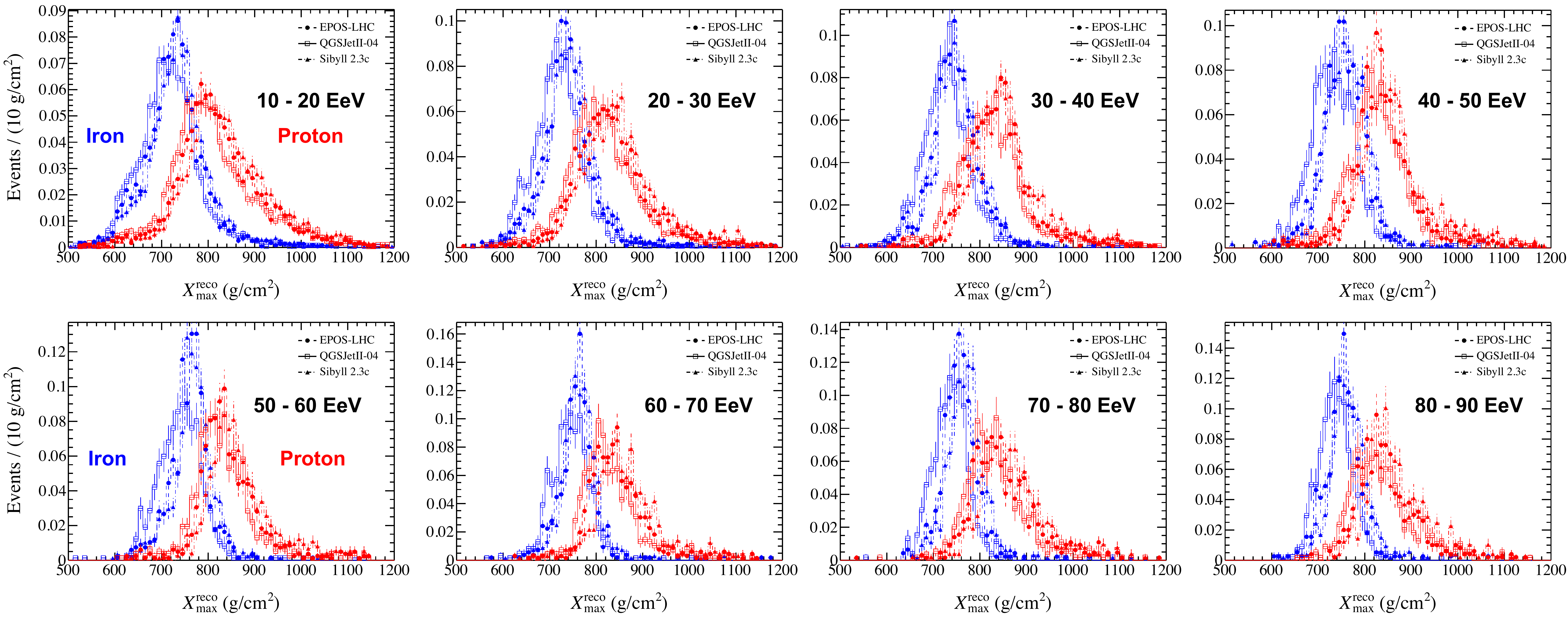}}
  \caption{Reconstruction bias on (a) $\langle X_{\max} \rangle$ and (b) $\sigma(X_{\max})$ evaluated by only the neural network first-guess estimation. (c) Reconstructed $X_{\max}$ distributions in each energy bin.}
  \label{bib:xmax}
\end{figure}

\section{Neural network first-guess estimation}
\begin{figure}[t]
  \centering
  \subfigure[3-fold Trigger efficiency]{\includegraphics[width=.49\linewidth]{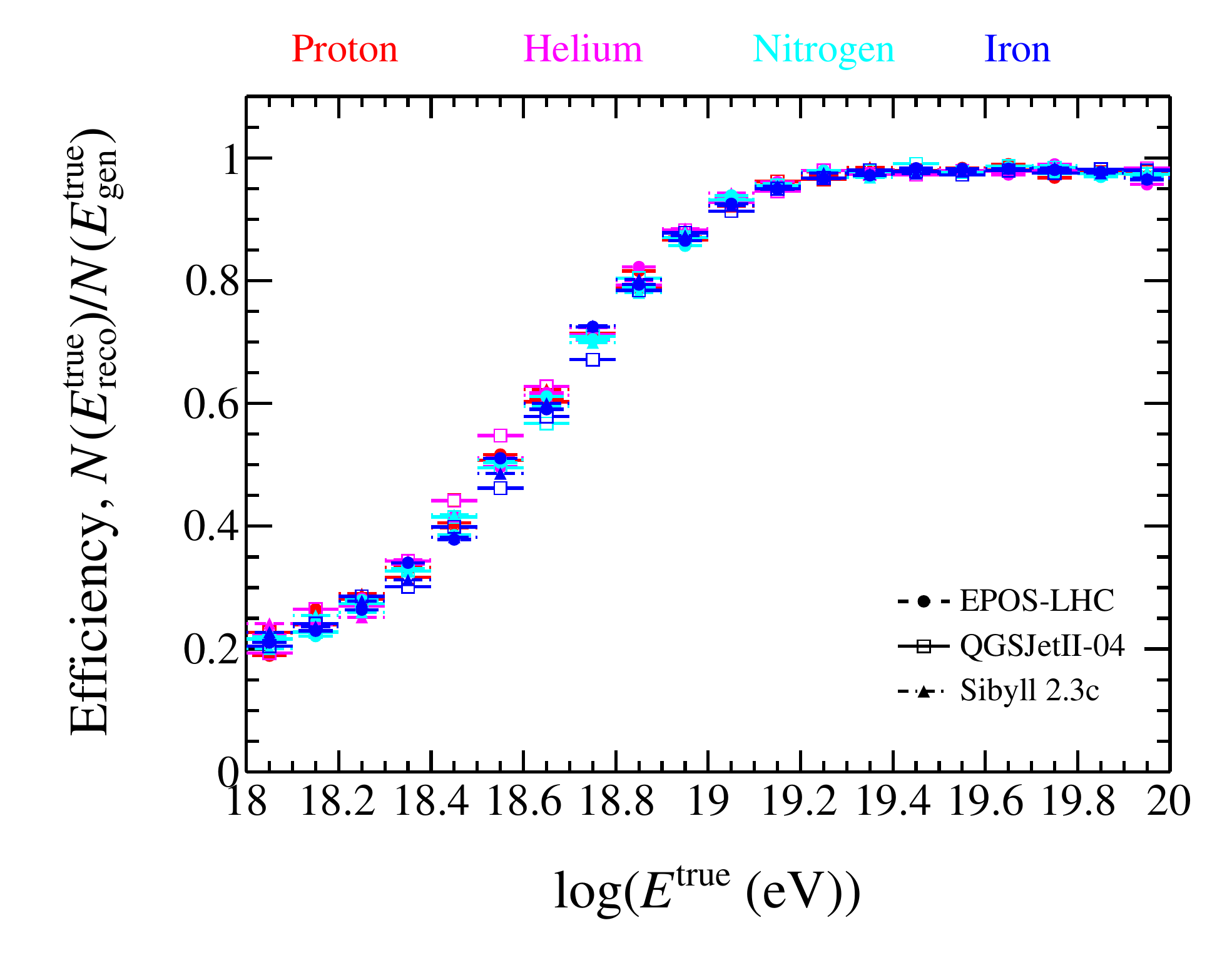}}
  \subfigure[Expected sensitivity of full FAST array]{\includegraphics[width=.49\linewidth]{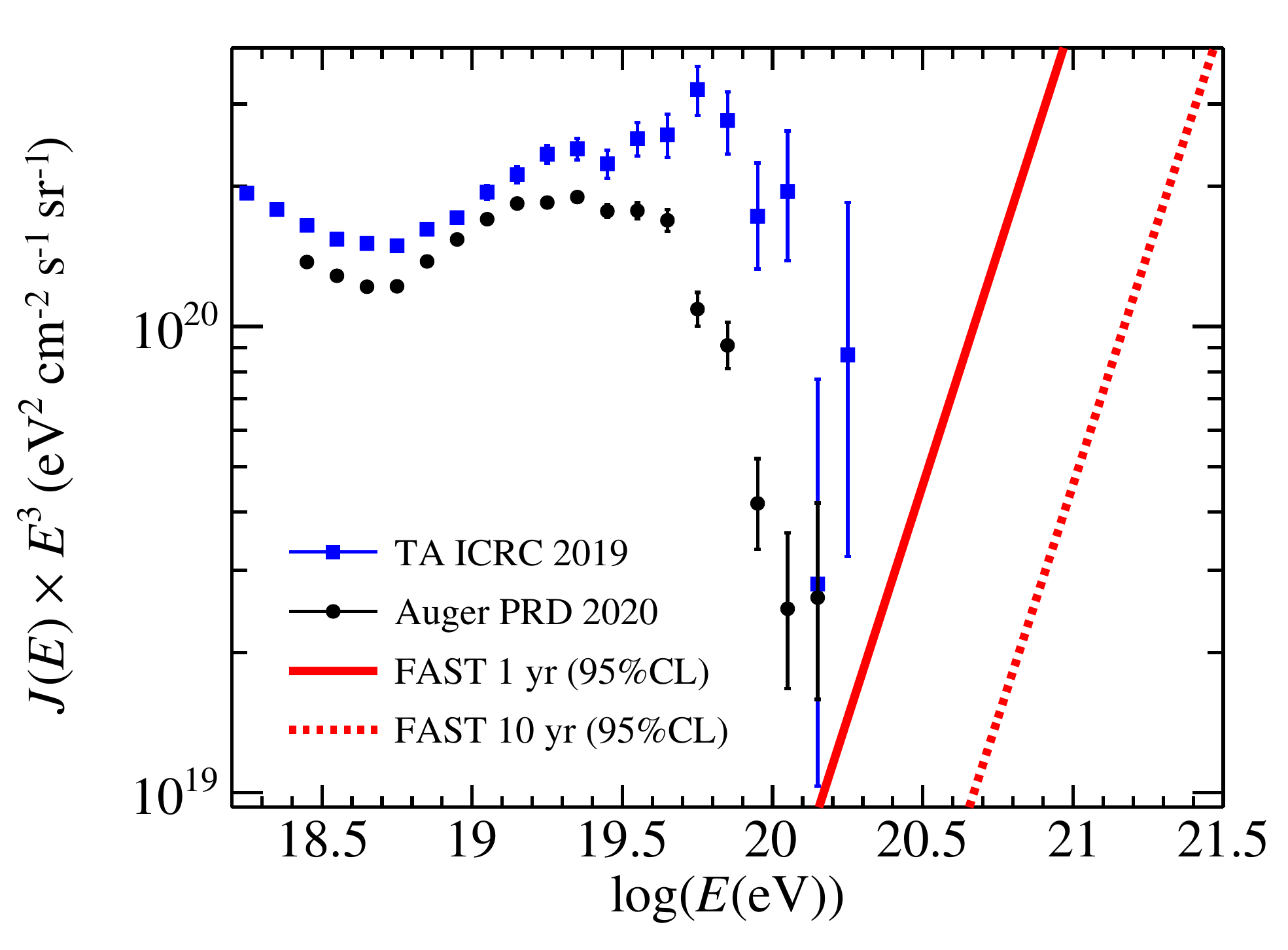}}
  \caption{(a) Trigger efficiency for 3-fold detections with a hypothetical FAST array. (b) Expected 95\% C.L. detectable sensitivities of the energy spectrum with the full FAST array of 500 stations compared to the spectra reported from TA~\cite{Ivanov:2020rqn} and Auger~\cite{Aab:2020rhr}.}
  \label{bib:eff_spec}
\end{figure}
The top-down reconstruction requires a reasonable first-guess geometry to reduce computational time. 
This is provided by a neural network first-guess estimation~\cite{Justin:2021phd}.
The total signal, centroid time, and pulse height of each PMT with a significant signal are used as inputs.
The outputs are six parameters: $X_{\max}$, energy, zenith, azimuth, and west-east/south-north core positions. The model uses the Keras/Tensorflow library with two hidden fully-connected layers.

The resolution and detection bias on $X_{\max}$ are evaluated by only applying this first-guess estimation for EASs of four primaries (proton, helium, nitrogen, and iron) with three hadronic interaction models (EPOS-LHC, QGSJetII-04 and Sibyll 2.3c)~\cite{bib:corsika}.
The EASs are generated with uniformly-distributed arrival directions and core positions randomly generated in the triangular array's inner circle.
The resolutions are 4.2 degrees in arrival direction, 465\,m in core position, 8\% in energy, and 30\,g/cm$^2$ on $X_{\max}$ at 40\,EeV for 3-fold coincidences without any quality cuts.
Figure~\ref{bib:xmax} shows a preliminary detection bias on $\langle X_{\max} \rangle$ and $\sigma(X_{\max})$, and also reconstructed $X_{\max}$ distributions in each energy bin.
Note that this performance is evaluated by only the neural network first-guess estimation. 
The full-chain performance of both top-down reconstruction and neural network first-guess estimation is being investigated.

The trigger efficiency for 3-fold detections is shown in Figure~\ref{bib:eff_spec}(a).
The FAST array has a 100\% efficiency above 20\,EeV. The energy threshold is related to the bias on the average $\langle X_{\max} \rangle$ and $\sigma(X_{\max})$ as shown in Figure~\ref{bib:xmax}.
Figure~\ref{bib:eff_spec}(b) is the expected sensitivity on the energy spectrum with a full-size FAST array.
We use an effective exposure of 90,000\,km$^2$\,sr per year to estimate our detectable flux at the 95\% confidence level. A full-sized FAST array will extend UHECR measurements beyond 300\,EeV.

\section{Developments for stand-alone observation of FAST array}
\begin{figure}
  \centering
  \includegraphics[width=1.0\linewidth]{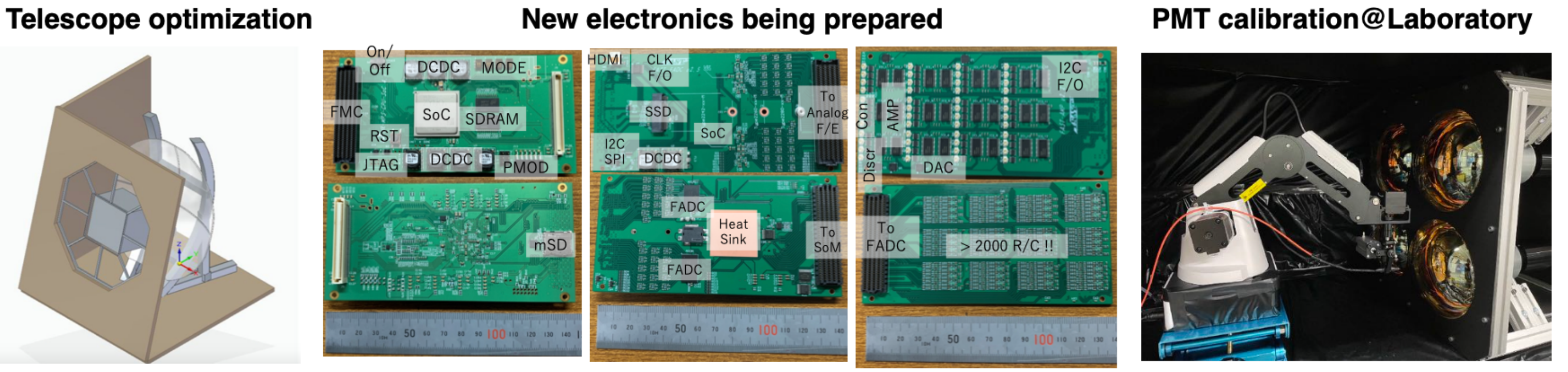}
  \caption{Developments for future stand-alone operations: the lighter telescope frame and mirror design with fewer pieces, the new electronics under development, and the PMT calibration system using a robotic arm.}
  \label{bib:next}
\end{figure}
Since these tests, several advances have been made: improvements in our telescope design, development of electronics with low-power consumption, and improvements in PMT calibration systems, as shown in Figure~\ref{bib:next}.  The improved electronics is particularly important as previous tests have capitalized on the infrastructure of existing FD detectors. These new electronics will allow for the first deployment of an independent, solar-powered FAST station, as well as permit stand-alone observation with the FAST array, an important step in validating our design and testing our expected resolution. The potential infield calibration could be performed using an extended uniform light source such as the integrating sphere~\cite{Vacula:2021}.

\section{Summary}
We have developed a low-cost, easily-deployed fluorescence detector optimized for detection of the highest energy cosmic rays in anticipation of a future array with 30,000\,km$^{2}$ of effective coverage. 
Three FAST prototypes have been installed at the Telescope Array Experiment, and one prototype has been installed at the Pierre Auger Observatory.  We have begun observations in both hemispheres and have demonstrated the viability of sophisticated, novel reconstruction methods. We will continue the steady operation of all four FAST prototypes and developments stand-alone observations with the FAST array. 

\section*{Acknowledgements}
This work was supported by JSPS KAKENHI Grant Number 21H04470, 20H05852, 18KK0381, 18H01225, 15H05443, 
and a research fund from the Hakubi Center for Advanced Research, Kyoto University.
This work was partially carried out by the joint research program of
the Institute for Cosmic Ray Research (ICRR) at the University of Tokyo.
This work was supported in part by NSF grant PHY-1713764, PHY-1412261 and by the Kavli
Institute for Cosmological Physics at the University of Chicago through
grant NSF PHY-1125897 and an endowment from the Kavli Foundation and its founder Fred Kavli.
The Czech authors gratefully acknowledge the support of the Ministry of Education,
Youth and Sports of the Czech Republic project No. LTAUSA17078, CZ.02.1.01/0.0/17\_049/0008422, LTT 18004. The Czech authors gratefully acknowledge the support of the Czech Academy of Sciences and Japan Society for the Promotion of Science within the bilateral joint research project with Kyoto University (Mobility Plus project JSPS 21-10).
The Australian authors acknowledge the support of the Australian Research Council, through Discovery Project DP150101622.
The authors thank the Pierre Auger and Telescope Array Collaborations for providing logistic support,  part of the instrumentation to perform the FAST prototype measurement, and productive discussions.

\fontsize{9pt}{0cm}\selectfont
\bibliography{fast_icrc2021}
\bibliographystyle{JHEP}

\end{document}